\documentclass{article}
\usepackage{spconf,amsmath,graphicx}
\usepackage{amssymb}
\usepackage{url}
\usepackage{cite}
\usepackage{booktabs}
\usepackage{flushend}

\title{LIWhiz: A Non-Intrusive Lyric Intelligibility Prediction System for the Cadenza Challenge}

\name{Ram C. M. C. Shekar and Iván López-Espejo$^1$\thanks{This work was funded by the Spanish Ministry of Science and Innovation under the ``Ramón y Cajal'' programme (RYC2022-036755-I).}}
\address{$^1$Department of Signal Theory, Telematics and Communications, University of Granada, Spain \\
\texttt{ramcharanmc@gmail.com, iloes@ugr.es}}

\begin{document}
%
\maketitle
\begin{abstract}
We present LIWhiz, a non-intrusive lyric intelligibility prediction system submitted to the ICASSP 2026 Cadenza Challenge. LIWhiz leverages Whisper for robust feature extraction and a trainable back-end for score prediction. Tested on the Cadenza Lyric Intelligibility Prediction (CLIP) evaluation set, LIWhiz achieves a root mean square error (RMSE) of 27.07\%, a 22.4\% relative RMSE reduction over the STOI-based baseline, yielding a substantial improvement in normalized cross-correlation.
\end{abstract}
%
\section{Introduction}
\label{sec:intro}

Understanding lyrics is crucial for music enjoyment, yet listeners with hearing loss often struggle to comprehend them clearly and effortlessly \cite{Alinka20}. Inspired by advances in speech intelligibility prediction (SIP), the development of lyric intelligibility prediction (LIP) methods \cite{Sharma19}---a largely unexplored area---could drive new lyric enhancement technologies. Such advances may not only improve music accessibility for listeners with hearing loss but also support general health.

Motivated by these considerations, we submit LIWhiz (\underline{L}yric \underline{I}ntelligibility \underline{Whiz}), a \emph{non-intrusive} LIP system, to the ICASSP 2026 Cadenza Challenge \cite{roa2026cadenza_overview}. LIWhiz builds on our previous work \cite{Haolan}, where we developed no-reference SIP models using a wav2vec 2.0 backbone \cite{w2v2} adapted for automatic speech recognition (ASR) under additive noise. In contrast to \cite{Haolan}, LIWhiz employs Whisper \cite{Whisper} for feature extraction. This choice is driven by its proven effectiveness in LyricWhiz \cite{LyricWhiz}, a training-free state-of-the-art automatic lyric transcription system that leverages Whisper to robustly recognize singing vocals.

\section{System Description}
\label{sec:system}

\begin{figure*}
    \centering
    \includegraphics[width=\linewidth]{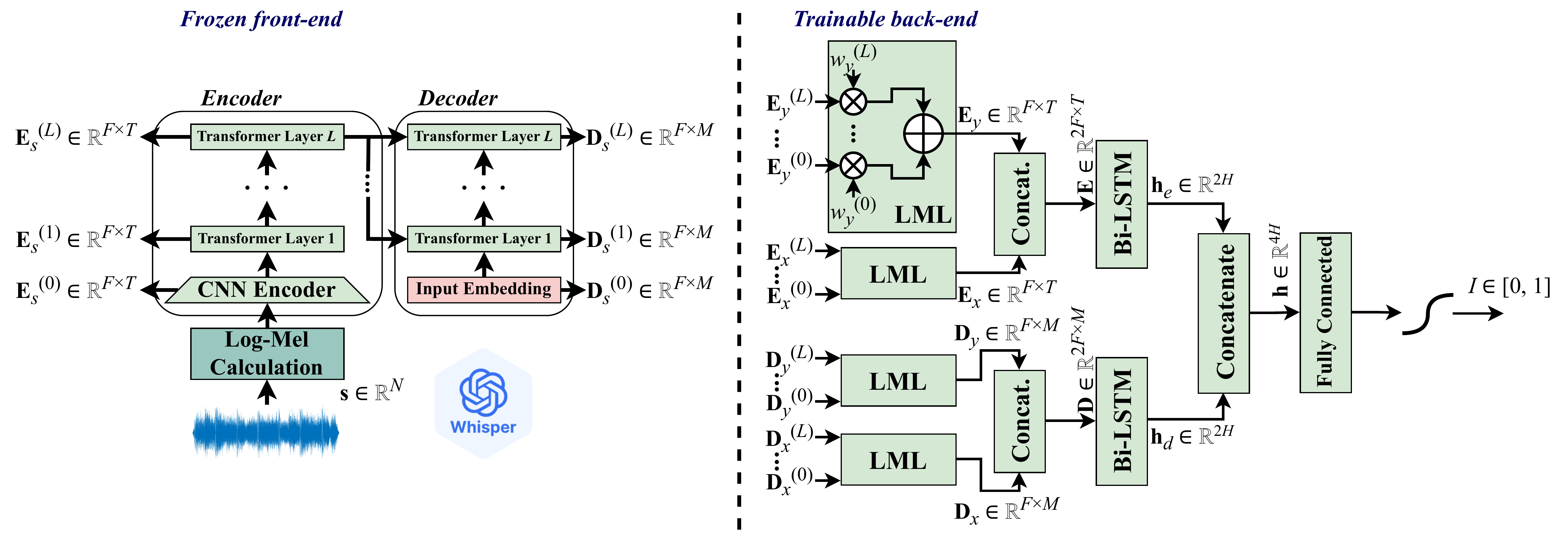}
    \caption{Diagram of the proposed LIP system, LIWhiz. The feature extractor (front-end), based on a frozen Whisper model, is shown on the left, where $\mathbf{s}\in\{\mathbf{x},\mathbf{y}\}$. The trainable back-end, which produces lyric intelligibility scores $I$ from features extracted from the original song excerpt $\mathbf{x}$ and its hearing-loss-simulated version $\mathbf{y}$, is shown on the right. LML denotes a linear mixing layer.}
    \label{fig:diagram}
\end{figure*}

A block diagram of LIWhiz is shown in Fig.~\ref{fig:diagram}. The front-end uses a frozen Whisper Large v3 model~\cite{Whisper} accessed via the Hugging Face Transformers library. For an input song excerpt $\mathbf{x}\in\mathbb{R}^N$ and its hearing-loss-simulated counterpart $\mathbf{y}\in\mathbb{R}^N$, the front-end independently extracts 66 feature maps: one from each of the $L=32$ encoder transformer layers plus the initial convolutional neural network (CNN) encoder block, and one from each of the $L=32$ decoder transformer layers plus the decoder input embedding. Each encoder (decoder) feature map is an $F\times T$ ($F\times M$) matrix, denoted $\mathbf{E}_x^{(l)}$, $\mathbf{E}_y^{(l)}$ ($\mathbf{D}_x^{(l)}$, $\mathbf{D}_y^{(l)}$), with $l=0,\dots,L$. Here, $F=1,280$ is the feature dimension, $T$ is the number of encoder time frames, and $M$ is the number of input tokens.

The trainable back-end predicts lyric intelligibility scores from $\mathbf{x}$ and $\mathbf{y}$. It begins with two linear mixing layers (LMLs) that fuse the Whisper encoder representations as $\mathbf{E}_x=\sum_{l=0}^L w_x^{(l)}\mathbf{E}_x^{(l)}$ and $\mathbf{E}_y=\sum_{l=0}^L w_y^{(l)}\mathbf{E}_y^{(l)}$, where $\{w_x^{(l)};\;l=0,\dots,L\}$ and $\{w_y^{(l)};\;l=0,\dots,L\}$ are learnable weights. The resulting embeddings $\mathbf{E}_x$ and $\mathbf{E}_y$ are concatenated into $\mathbf{E} \in \mathbb{R}^{2F \times T}$ and fed into a Bi-LSTM. The Bi-LSTM's final hidden state $\mathbf{h}_e \in \mathbb{R}^{2H}$ ($H=512$) is concatenated with $\mathbf{h}_d \in \mathbb{R}^{2H}$, obtained in parallel from the decoder feature maps (see Fig.~\ref{fig:diagram}). This vector, $\mathbf{h}\in\mathbb{R}^{4H}$, is passed through a single-neuron fully-connected layer with sigmoid activation to produce the lyric intelligibility score $I \in [0,1]$.

We hypothesize that including the original song excerpt $\mathbf{x}$ alongside its hearing-loss-simulated version $\mathbf{y}$ provides LIWhiz with cues to better adapt to the degree of hearing loss, yielding more accurate lyric intelligibility predictions.

\section{Experimental Setup and Results}
\label{sec:res}

To train the back-end, we use only the training partition of the Cadenza Lyric Intelligibility Prediction (CLIP) dataset \cite{roa2025clip}, which contains thousands of audio excerpts of unfamiliar Western popular music paired with listening-test lyric intelligibility scores. Ground-truth lyric transcripts are also available but not used, making our system fully non-intrusive.

Prior to training and inference, stereo audio from the CLIP dataset is converted to mono and downsampled to 16~kHz to ensure compatibility with Whisper. Since lyric intelligibility scores are not initially provided for the CLIP validation set, we perform $k$-fold cross-validation with $k=10$ during training. Early stopping with a patience of 10~epochs is used for regularization, and training runs on an NVIDIA V100 Tensor Core GPU for a maximum of 30~epochs with the AdamW optimizer (learning rate of $10^{-3}$). Given that the primary evaluation metric is root mean square error (RMSE), this is also used as the loss function. During inference, the final score is obtained by averaging the predictions from the $k=10$ models resulting from cross-validation.

\begin{table}[t]
\centering
\caption{RMSE (\%) and NCC results on the CLIP validation and evaluation sets for LIWhiz, alongside non-intrusive STOI- and intrusive Whisper-based baselines \cite{roa2026cadenza_overview}.}
\label{tab:results}
\resizebox{\columnwidth}{!}{%
\begin{tabular}{l|cc|cc}
\toprule
\textbf{System} & \multicolumn{2}{c|}{\emph{Validation set}} & \multicolumn{2}{c}{\emph{Evaluation set}} \\
\cmidrule(r){2-3} \cmidrule(r){4-5}
 & \textbf{RMSE (\%) $\downarrow$} & \textbf{NCC $\uparrow$} & \textbf{RMSE (\%) $\downarrow$} & \textbf{NCC $\uparrow$} \\
\midrule
Baseline STOI & 36.11 & 0.14 & 34.89 & 0.21 \\
Baseline Whisper & 29.32 & 0.59 & 29.08 & 0.58 \\
\midrule
LIWhiz & \textbf{27.13} & \textbf{0.67} & \textbf{27.07} & \textbf{0.65} \\
\bottomrule
\end{tabular}}
\end{table}

Table~\ref{tab:results} reports RMSE (in percentages) and normalized cross-correlation (NCC) results on the CLIP validation and evaluation sets. LIWhiz is compared with the non-intrusive STOI- and intrusive Whisper-based baselines provided by the challenge organizers \cite{roa2026cadenza_overview}. As shown, LIWhiz substantially outperforms the baselines on both metrics and sets.

\begin{figure}[t]
	\centering
	\begin{minipage}{0.5\linewidth}
		\includegraphics[width=\linewidth]{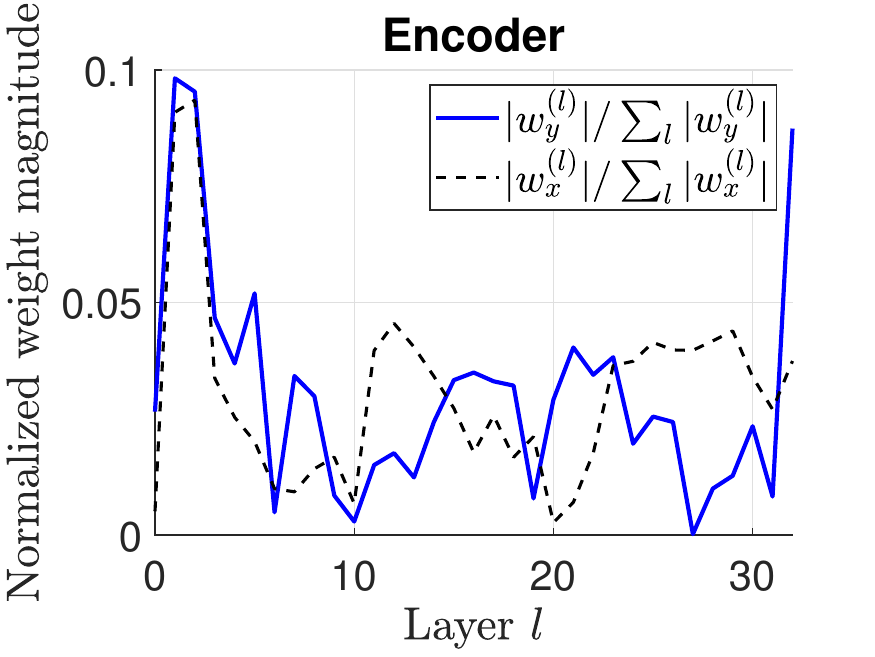}
	\end{minipage}\hfill
	\begin{minipage}{0.5\linewidth}
		\includegraphics[width=\linewidth]{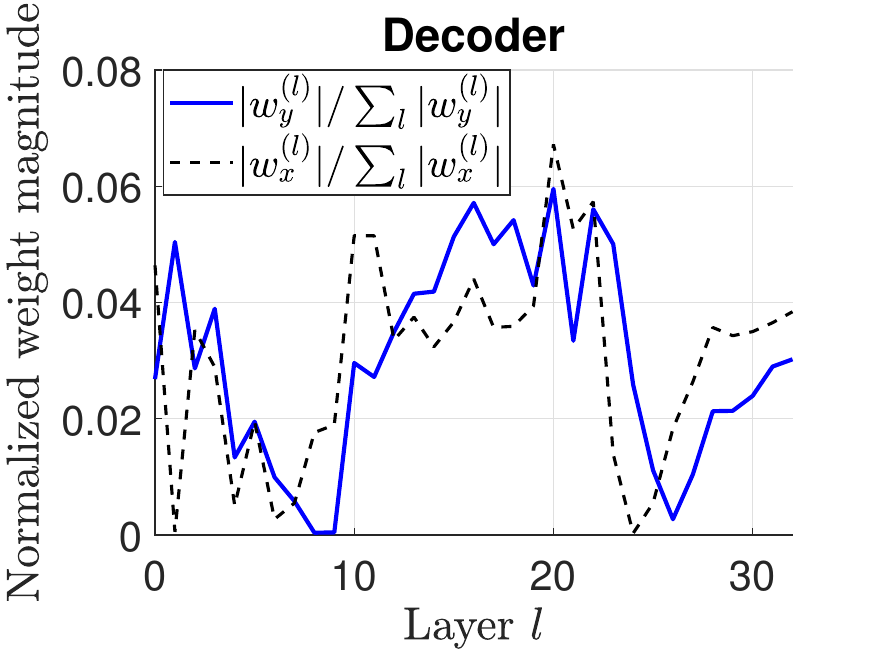}
	\end{minipage}
	\caption{Normalized absolute values of the learned LML weights for both the encoder (left) and decoder (right).}
	\label{fig:weights}
\end{figure}

Fig. \ref{fig:weights} displays the normalized absolute values of the learned LML weights for both the encoder and decoder. The encoder weights indicate that both the original ($\mathbf{x}$) and hearing-loss-simulated ($\mathbf{y}$) excerpts rely predominantly on the shallowest layers, consistent with low-level acoustic encoding, while the deepest layer carries relatively more weight for the simulated signal. In the decoder, LIWhiz focuses on the middle layers, where representations have already integrated information from the encoder, producing features most predictive of lyric intelligibility.

Finally, as hypothesized in Section~\ref{sec:system}, including the original song excerpt $\mathbf{x}$ alongside $\mathbf{y}$ leads to a slight improvement in LIP performance. When the $\mathbf{x}$ branches in Fig.~\ref{fig:diagram} are removed, the validation and evaluation RMSE increase from 27.13\% and 27.07\% (see Table \ref{tab:results}) to 27.36\% and 27.34\%, respectively.

\section{Conclusion}
\label{sec:conclusion}

In this work, we introduced LIWhiz, a non-intrusive LIP system inspired by SIP and powered by Whisper-based feature extraction, motivated by Whisper's state-of-the-art performance in automatic lyric transcription---an intrinsically related task. Experimental results demonstrate that LIWhiz substantially outperforms the non-intrusive STOI- and intrusive Whisper-based baselines provided by the challenge organizers.

\bibliographystyle{IEEEbib}
\bibliography{strings,refs}

\end{document}